# The Statistical Analyses of Flares Detected In B Band Photometry of UV Ceti Type Stars

**H. A. Dal, S. Evren**

*Ege University, Science Faculty, Department of Astronomy and Space Sciences, 35100 Bornova, İzmir, Turkey*

**Abstract**
In this study, we present the unpublished flare data collected from 222 flares detected in B band observations of five stars and the results derived by the statistical analysis and modelling of these data. Using a statistical analysis method, six basic properties have been found from all models and analyses for the flares detected in B band observation of UV Ceti type stars. We have also compare U and B bands in respect to the analysis results. This gave us a chance to test whether the methods using in the analyses are statistically trustworthy, or not. According to the analyses, 1) The flares were separated into two types as fast and slow flares. 2) There is 16.2±3.7 times difference between the mean averages of the equivalent durations of the slow and the fast flares. 3) It is no matter how long the flare total durations, maximum flare energy can reach a different *Plateau* level of energy for each star. 4) The Plateau values of EV Lac and EQ Peg are absolutely higher than the others. 5) Minimum values of the flare total durations increase toward the later spectral types. This value is called Half-Life value in models. 6) Both maximum flare rise times and flare total durations obtained from the observed flares are decreasing toward the later spectral type.

**1. Introduction**
The first flare was detected on the Sun by R. C. Carrington and R. Hodgson on 1 September, 1859 (Carrington, 1859; Hodgson, 1859), while it was detected from UV Ceti stars in 1939 (Van Maanen, 1940). The observations demonstrate that the flare activity is a common property among late-type stars. Although several stars between the spectral type F and M exhibit flare activity, but the stars exhibiting the flare activity are generally dMe stars (Audard et al., 2000; Gershberg, 2005). According to Rodonó (1986), the incidence of red dwarfs in our Galaxy is 65%, and seventy-five percent of them show the flare activity. The flare stars known as UV Ceti stars are assumed as the young stars. They are generally pre-main sequence stars (Mirzoyan, 1990; Gershberg, 2005; Benz & Güdel, 2010). However, there are some stars in the close binary system, and they also exhibit the flare activity although they are not as young as the pre-main sequence stars (Poveda et al., 1996a,b; Rocha-Pinto et al., 2002). The late-type young stars exhibit the flare activity from the radio to the X-Ray bands (Osten et al., 2005; Crespo-Chacón et al., 2007).

Discovering some stars with high flare frequency such as UV Cet, YZ CMi, EV Lac, AD Leo and EQ Peg, both the number of known flares and the variety of flare light variations increased (Moffett, 1974; Gershberg, 2005). Although there are lots of light variation shapes for the flares, there are two main types of the shape for flare light variations (Dal & Evren, 2010a). One of them is the fast flares and the other is slow flares. The terms of fast and slow flares were used for the first time by Haro & Parsamian (1969). According to Haro & Parsamian (1969), if the rise time of a flare is higher than 30 minutes, the flare is a slow flare. If the rise time is below 30 minutes, it is a fast flare. Kunkel stated an idea in his PhD thesis in 1967 (Kunkel, 1967) that the shape of the flare light variations is caused by combinations of some fast and slow flares. Considering the light variations of observed flares, Osawa et al. (1968) identified two flare types. On the other hand, Oskanian (1969) separated flares into four types. Depending on just flare light variations, Moffett (1974) classified flares such as classic, complex, slow flares or flare events. Gurzadian (1988), who modelled the white light flares, separated the flares detected in UV Ceti stars into two types such as fast and slow flares. Gurzadian (1988) asserted that the slow flares are produced by thermal processes and they constitute ninety-five per cent of all flares observed in UV Ceti type stars. All other flares are fast flares produced by non-thermal processes. According to Gurzadian (1988), there is large energy difference between these two type flares. Dal & Evren (2010a) examined the distributions of flare equivalent durations versus flare rise times. They demonstrated that the value of 3.5 in the ratio of flare decay time to flare rise time is a limit value between fast and slow flares. According to the value of 3.5, the flares are separated as fast and slow flares. Using a statistical analysis method, Dal & Evren (2010a) computed the difference between the mean averages of the equivalent durations of the slow and the fast flares. The mean averages of the equivalent durations of the fast flares was found to be 157 times higher than that of the slow flares.

The flare activity caused by magnetic activity in dMe stars has not been fully explained yet (Benz & Güdel, 2010). It is seen that there is large energy difference between the flares observed in different type stars. Although there is large difference between the energies of the flares detected on the Sun and dMe stars, the flare activity seen on the dMe stars has been tried to explain by the processes in the Solar Flare Event. The energy source of flare events is explained by magnetic reconnection processes (Benz & Güdel, 2010; Gershberg, 2005; Hudson & Khan, 1997). To understand properly all process of flare event for dMe stars, first of all, it could be useful to find out all similarities and differences between flare behaviours by examining the phenomenon star-to-star. For instance, looking over the flare energy spectra could be useful. This distribution can show how flare energies range from one star to the other. In this respect, the flare energy spectra of UV Ceti type stars were examined in many studies such as Gershberg (1972); Lacy et al. (1976); Walker (1981); Gershberg & Shakhovskaya (1983); Pettersen et al. (1984) and, Mavridis & Avgoloupis (1986). For instance, the flare energy spectra were studied by Gershberg (1972) for AD Leo, EV Lac, UV Cet and YZ CMi. In another study, the flare energy spectra for lots of stars in galactic field were compared with flare energy spectra of some stars from Pleiades cluster and Orion association by Gershberg & Shakhovskaya (1983). The power law exponent of the flare energy spectra for the stars from Orion association is higher than all the others. Pleiades stars are located just below them. The power law exponent of the flare energy spectra for the stars from galactic field is located at bottom. It seems that these differences between the power law exponents of the flare energy spectrum for the stars from different sources are due to age. The same separation is seen among some field stars, too. Moreover, the flare frequencies were computed to investigation for flare activity in some other studies (Pettersen et al., 1983; Ishida et al., 1991; Leto et al., 1997). Two flare frequencies were generally computed in these studies. Using these frequencies, the behaviour of the flare activity was examined for each star.

The flare activity of AD Leo was discovered for the first time by Gordon & Kron (1949). Crespo-Chac´on et al. (2006) computed flare frequency for AD Leo as 0.71 $h^{-1}$. Ishida et al. (1991) examined the variations of flare frequencies for AD Leo and demonstrated that the frequencies were not varying. EV Lac is the other star in our programme. It has known from 1950 that EV Lac exhibits flare activity (Lippincott, 1952; Van de Kamp, 1953). Comparing the variation of the flare frequency with the seasonal averages of B band magnitude, (Mavridis & Avgoloupis, 1986) found that the activity cycle is about 5 years for EV Lac. On the other hand, Ishida et al. (1991) demonstrated that there was no flare frequency variation from 1971 to 1988. In another study, Leto et al. (1997) showed that flare frequencies of EV Lac increased from 1968 to 1977. EQ Peg is another active flare star in the programme, whose flare activity was discovered by Roques (1954). EQ Peg is known as a metal-rich star (Fleming et al., 1995). EQ Peg is a visual binary (Wilson, 1954). Both of the components are flare stars (Pettersen et al., 1983). Angular distance between the components is given as between 3".5 and 5".2 (Haisch et al., 1987; Robrade et al., 2004). One of the components is of 10.4 mag and the other is of 12.6 mag in V band (Kukarin, 1969). Observations show that flares of EQ Peg generally come from the fainter component (Fossi et al., 1995). Rodon´o (1978) proved that 65% of the flares come from the faint component and about 35% from the brighter component. The fourth star in this study is V1054 Oph, whose flare activity was discovered by Eggen (1965). V1054 Oph (= Wolf 630ABab, Gliese 644ABab) is a member of Wolf star group (Joy, 1947; Joy & Abt, 1974; Mazeh et al., 2001). Wolf 630ABab, Wolf 629AB (= Gliese 643AB) and VB8 (= Gliese 644C), are the main members of a triplet system. Wolf 630 and Wolf 629 are visual binaries and they are separated 72" from each other. Wolf 630AB is a close visual binary in itself and there is an angular distance about 0".218 between A and B components. Wolf 629AB is a spectroscopic binary. B component of Wolf 629AB also seems to be a spectroscopic binary. VB8 is 220" far away from the other two main components (Joy, 1947; Joy & Abt, 1974). The fifth star is a young disk star, V1005 Ori (Veeder, 1974). Its first flare was obtained by Shakhovskaya (1974).

In this study, the results obtained from the analyses of 222 flares detected in B band observations of five UV Ceti type stars with high flare frequencies. The types of the flares detected in B band were examined by the method described by Dal & Evren (2010a) and the results were compared with the analyses of the flares detected in U band presented by Dal & Evren (2010a). In addition, the distributions of flare equivalent durations versus flare total durations were obtained for each star to compare the behaviours of flare activity among programme stars. Using the method described by Dal & Evren (2010b) for U band flares, all the distributions were modelled by the *One Phase Exponential Association* (hereafter *OPEA*) (Motulsky, 2007; Spanier & Oldham, 1987) function. The results obtained from the models were tested by a statistical analysis method, *Independent Samples t-Test* (hereafter *t-Test*) (Wall & Jenkins, 2003; Dawson & Trapp, 2004; Motulsky, 2007). Finally, the flare frequencies computed for each star are presented for each observing season. It is discussed which flare parameter is the best indicator for the flare activity level. Analysing B band flares with using the method described by Dal & Evren (2010a,b), we have tested whether these methods are statistically trustworthy, or not.

**Table 1:** Basic parameters for the target studied and its comparison (C1) and check (C2) stars. Columns list: Star name; standard V mag and B-V colours for quiet phase of them.

| Stars | V (mag) | B-V (mag) | Spectral Type | Distance (pc) | Age |
|---|---|---|---|---|---|
| **AD Leo** | 9.388 | 1.498 | M3 | 4.90 | ~ 200$^a$ Myr |
| C1: HD 89772 | 8.967 | 1.246 | K6 K7 | - | - |
| C2: HD 89471 | 7.778 | 1.342 | K8 | - | - |
| **EV Lac** | 10.313 | 1.554 | M3 | 5.00 | ~ 300$^b$ Myr |
| C1: HD 215576 | 9.227 | 1.197 | K6 | - | - |
| C2: HD 215488 | 10.037 | 0.881 | K1 K2 | - | - |
| **EQ Peg** | 10.170 | 1.574 | M3-M4 | 6.58 | YD$^c$ |
| C1: SAO 108666 | 9.598 | 0.745 | G8 | - | - |
| C2: SAO 91312 | 9.050 | 1.040 | K3-K4 | - | - |
| **V1005 Ori** | 10.090 | 1.307 | K7 | 26.70 | ~ 35$^d$ Myr |
| C1: BD +01 870 | 8.800 | 1.162 | K5 | - | - |
| C2: HD 31452 | 9.990 | 0.920 | K2 | - | - |
| **V1054 Oph** | 8.996 | 1.552 | M3 | 5.70 | ~ 5$^e$ Gyr |
| C1: HD 152678 | 7.976 | 1.549 | M3 | - | - |
| C2: SAO 141448 | 9.978 | 0.805 | K0 | - | - |

$^a$ A member of the Castor Moving Group (Montes et al., 2001).
$^b$ A member of the Ursa Major Group (Montes et al., 2001).
$^c$ A member of the young disk population in the galaxy (Veeder, 1974).
$^d$ A member of the IC 2391 Supercluster (Montes et al., 2001).
$^e$ (Mazeh et al., 2001).

**2. Observations**

The observations were acquired with a High Speed Three Channel Photometer attached to a 48 cm Cassegrain type telescope at Ege University Observatory. Using a tracking star set in a second channel of the photometer, the observations were continued in standard Johnson B band with the exposure times between 1 and 7 s. The basic parameters of all programme stars and their comparisons are given in Table 1. The parameters such as star name, magnitude in V band, B-V colour index, spectral types, distance (pc) and approximate age are listen in the table, respectively. Considering B-V colour indexes, the spectral types were taken from Tokunaga (2000). The distances of the stars were taken from Gershberg et al. (1999). Although the programme and comparison stars are so close on the sky, differential extinction corrections were applied. The extinction coefficients were obtained from the observations of the comparison stars on each night. Moreover, the comparison stars were observed with the standard stars in their vicinity and the reduced differential magnitudes, in the sense variable minus comparison, were transformed to the standard system using procedures outlined in Hardie (1962). The standard stars are listed in the catalogues of Landolt (1983) and Landolt (1992). Heliocentric corrections were applied to the times of the observations. The standard deviation of observation points acquired in standard Johnson B band is about $0^m.009$ on each night. Observational reports of all programme stars are given in Table 2. It is seen that there is no variation of differential magnitudes in the sense comparison minus check stars.

Table 2: Observational reports of the each programme star for each observing season.

| Stars | Observing Year | HJD Interval (+2400000) | Observing Duration (h) | Flare Number |
|---|---|---|---|---|
| AD Leo | 2004 - 2005 | 53377 - 53514 | 39.69083 | 36 |
|  | 2005 - 2006 | 53717 - 53831 | 38.91617 | 48 |
|  | 2006 - 2007 | 54048 - 54248 | 24.11472 | 9 |
| EV Lac | 2004 | 53202 - 53312 | 6.51111 | - |
|  | 2005 | 53554 - 53606 | 30.31083 | 22 |
|  | 2006 | 53940 - 53996 | 48.7625 | 30 |
| EQ Peg | 2004 | 53236 - 53335 | 9.50972 | - |
|  | 2005 | 53621 - 53686 | 37.46639 | 29 |
| V1005 Ori | 2004 - 2005 | 53353 - 53453 | 30.14028 | 5 |
|  | 2005 - 2006 | 53640 - 53812 | 31.40111 | 29 |
| V1054 Oph | 2004 | 53136 - 53196 | 18.4375 | - |
|  | 2005 | 53502 - 53564 | 33.16222 | 14 |

All the flare equivalent durations and energies were computed using Equations (1) and (2) taken from Gershberg (1972).

$$P_B = \int [(I_{flare} - I_0)/I_0] dt \quad \text{...................... (1)}$$

where $I_0$ is the intensity of the star in quiescent level; $I flare$ is the intensity of the star during flare.

$$E_B = P_B x L_B \quad \text{...................... (2)}$$

where $E_B$ is the flare energy in B band, and $P_B$ is the flare equivalent duration described by Equation (1). $L_B$ is the luminosity of star in quiescent level. In this study, we use the flare equivalent durations instead of the flare energy due to the luminosity term in Equation (2). The luminosities of stars from different spectral types are very different from each other. Even if the equivalent durations of two flares obtained from two stars in different spectral types were the same, the calculated energies of these flares would be different due to different luminosities of these spectral types. Therefore, we could not use these flare energies in the same analyses. However, flare equivalent duration depends only on the flare power. In addition, the given distances of the same star in different studies are quite different. The calculated luminosities became different due to these different distances.

Some samples of flares detected in B band observations of five stars in this study are shown in Figures 1 - 4. In the figures, filled circles represent intensities of the observation points, while dotted lines represent the quiescent level. It is more important to note that the brightness of a star without any flare or any sudden variation was taken as a quiescent level of the brightness of the star on each night. Considering this level, all flare parameters were calculated for each night. In addition, as seen from Figure 2, the flares have sometimes several peaks. In this case, the flare maximum times and amplitudes were calculated from the first highest peak. The flares shown in Figure 1 are two samples of the flares frequently detected in the observations of UV Ceti type stars. The shape of the light variation in Flare A resembles the light variation of the flares observed in hard X-Ray observations of the Sun. As the sample seen in Figure 2, some complex flares were detected in the observations. The flare shown in Figure 2 is a combination of four fast flares, Flare A, B, C and D. Actually, Flare B, C and D have also complex structures. In this study, the flares like these complex samples were not used in the analyses. Moreover, the light variations of some flares were not completed because we did not carry on observation until the flare completely decreased to quiescent phase due to sunrise. Flare D shown in Figure 2 could be a sample of uncompleted flares. We did not use these uncompleted flares in the analyses.

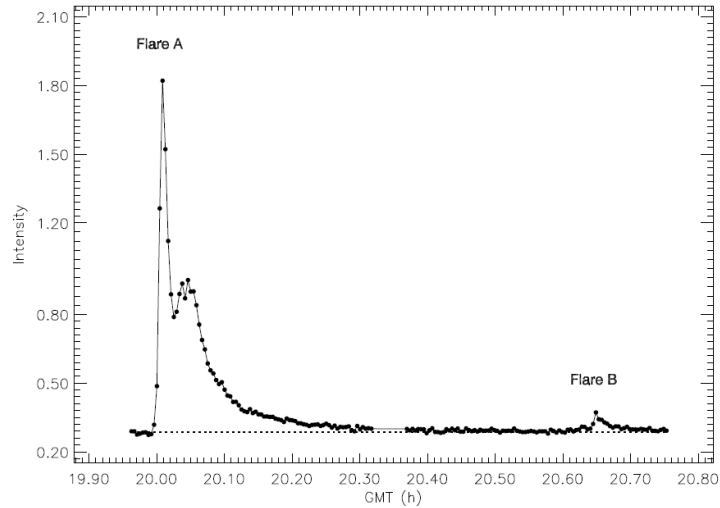

**Figure 1:** A fast flare detected in the observation of EQ Peg on 12 September, 2005.

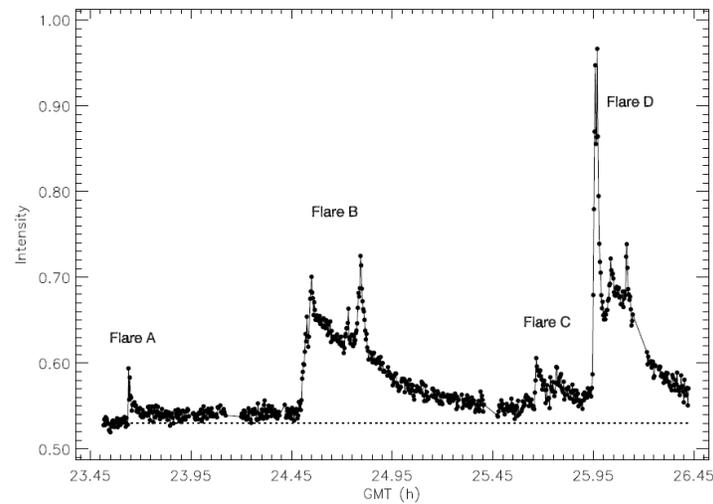

**Figure 2:** A combination of some fast flares detected in the observation of AD Leo on 10 February, 2005.

Some samples of the slow flares are shown in Figures 3 and 4. However, Flare A seen in Figure 4 is a combination of a fast and a slow flares.

For each flare detected in the observations, flare parameters such as flare rise time ($T_r$, s), flare decay time ($T_d$, s), flare total duration ($T_t$, s), which is the sum of both rise and decay times, flare equivalent duration ($P_B$, s) and flare energy ($E_B$, *ergs*), were computed. Considering the quiescent level computed for each night, all the computed parameters are listed in Table 3 (Table 3 is given as on-line form). In the columns of the table, star names, observing dates, HJDs of the flare peaks, flare rise times (s), flare total durations (s), flare equivalent durations (s), flare energies (*ergs*) and amplitudes of the flares (*mag*) are listed, respectively.

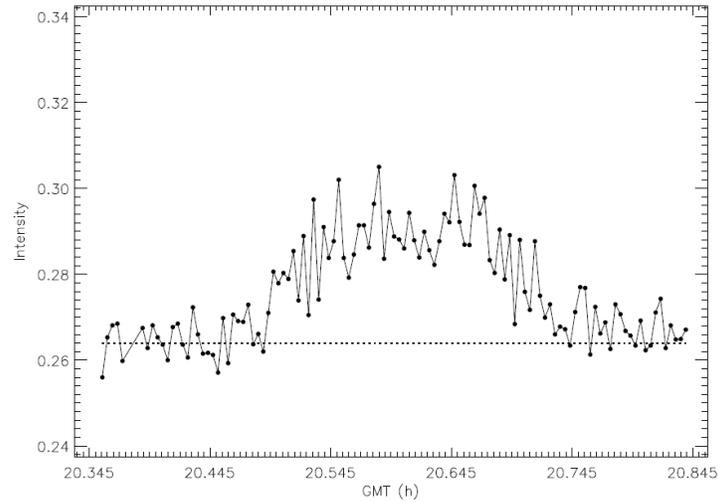

**Figure 3:** A slow flare detected in the observation of EV Lac on 8 August, 2006.

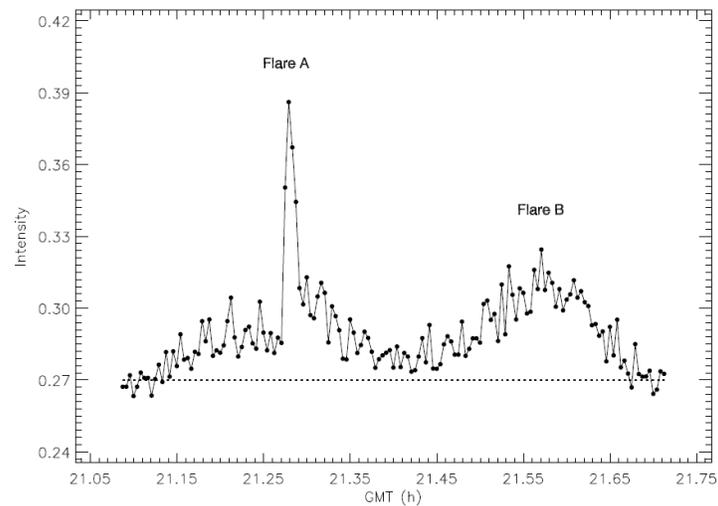

**Figure 4:** Some flares detected in the observations of EV Lac on 8 August, 2006. Flare A is a combination of a slow and a fast flares. The Flare B is a single slow flare.

**Table 3:** All the calculated parameters of observed flares are listed. From the first column to the last, star name, the date of observation, HJD of flare maximum moment, flare rise time (s), decay time (s), flare equivalent duration (s), flare energy (ergs), and flare amplitude (mag), respectively. In the last column, flare types are given. In the last column, the symbol "asterisk (*)" indicates that the flare was used in the analyses. The table seen below is a sample part of the whole table given as on-line form.

| Star Name | Observation Date | HJD Of Maximum (+ 24 00000) | Rise Time (s) | Decay Time (s) | Equivalent Duration (s) | Flare Energy (ergs) | Flare Amplitude (mag) | Flare Type |
|---|---|---|---|---|---|---|---|---|
| AD Leo | 06.01.05 | 53377.50656 | 24 | 24 | 0.61915 | 1.97754E+30 | 0.054 | Slow * |
| AD Leo | 06.01.05 | 53377.59767 | 12 | 540 | 6.62751 | 2.11681E+31 | 0.086 | Fast * |
| AD Leo | 10.01.05 | 53381.51295 | 252 | 672 | 20.90306 | 6.67639E+31 | 0.071 | Slow |
| AD Leo | 10.01.05 | 53381.52962 | 84 | 672 | 14.61875 | 4.66920E+31 | 0.045 | Fast |
| AD Leo | 10.02.05 | 53412.49015 | 300 | 804 | 55.86125 | 1.78420E+32 | 0.127 | Slow |
| AD Leo | 10.02.05 | 53412.52807 | 312 | 3636 | 424.29464 | 1.35519E+33 | 0.297 | Fast * |
| AD Leo | 10.02.05 | 53412.57460 | 384 | 828 | 78.55977 | 2.50918E+32 | 0.135 | Slow |
| AD Leo | 10.02.05 | 53412.58682 | 228 | 1668 | 362.66202 | 1.15833E+33 | 0.597 | Fast |
| AD Leo | 11.02.05 | 53413.46309 | 60 | 48 | 1.37084 | 4.37843E+30 | 0.035 | Slow * |

## 3. Fast and Slow Flares

It is expected that there must be a large energy difference between the slow flares produced by thermal processes and the fast flare produced by non-thermal processes (Gurzadian, 1988). The light variations of the stellar white-light flares have two main phases. One of them is the impulsive phase, in which large amounts of energy is suddenly released. The second one is the main phase, in which all the released energy is emitted to space (Gurzadian, 1988; Benz & Güdel, 2010). The sudden energy release in the impulsive phase of the fast flares can be seen with the hard X-Ray observations of the solar flares (Benz & Güdel, 2010; Wang et al., 2011). The main difference between the fast and slow flares is seen in the impulsive phase. In this phase, the rate of increase in the brightness of the fast flares is remarkable higher than the slow flares (Gurzadian, 1988; Gershberg, 2005). Although a sudden increasing is seen in the impulsive phase of the fast flares as seen from Flare A in Figure 1, the impulsive phases of slow flares is quite slow as seen from the flare in Figure 3. The light variation shapes of the fast flares are the same with the light variations seen in the hard X-Ray light curves of the solar flares (Coyner & Alexander, 2009). In addition, the fast or slow increasing is also exist in the light variations of solar $H\alpha$ flares (Temmer et al., 2001).

Consequently, if the energies of a fast and a slow flares, whose rise times are the same, are compared, it must be seen that there is a large difference between the energies of these two type flares. The light variation shapes of 222 flares were looked over. 129 flares were identified in 17 different rise times. 59 flares in these 17 different rise times have a rapid increasing in the impulsive phases. 77 flares have a slow increasing in the impulsive phase. Thus, both some fast and some slow flares were identified in each rise time of all 17 rise times. For instance, we identified 9 flares, whose rise time is 45 s. Two of them are the fast flares, while seven of them are the slow flares.

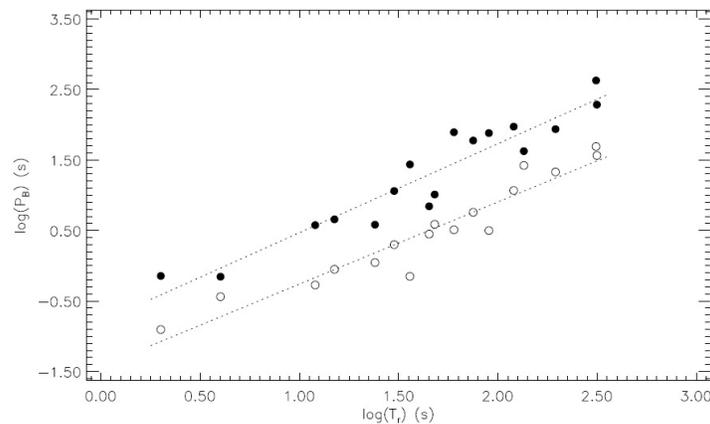

**Figure 5:** The distributions of the flare equivalent durations versus flare rise times are shown for 59 fast (filled circles) and 77 slow (open circles) flares. In the figure, dotted lines represent the linear fits.

**Table 4:** Using the Least-Squares Method and GrahpPad Prism V5.02 (Dawson & Trapp, 2004) software, the parameters of linear fits derived for the distributions of the flare equivalent durations versus flare rise times are listed for both fast and slow flares.

| Flare Type: | Fast Flare | Slow Flare |
|---|---|---|
| Slope: | 1.262±0.109 | 1.165±0.093 |
| y-intercept (when x = 0.0): | -0.793±0.192 | -1.424±0.163 |
| **95% Confidence Intervals** | | |
| Slope: | 1.029 to 1.495 | 0.967 to 1.363 |
| y-intercept (when x = 0.0): | -1.201 to -0.3846 | -1.770 to -1.077 |
| **Goodness of Fit** | | |
| $r^2$: | 0.899 | 0.913 |
| **Is slope significantly non-zero?** | | |
| F: | 133.2 | 157.4 |
| DFn, DFd: | 1.000, 15.00 | 1.000, 15.00 |
| p-value: | < 0.0001 | < 0.0001 |
| Deviation from zero? | Significant | Significant |
| **Runs test** | | |
| Points above line: | 8 | 10 |
| Points below line: | 9 | 7 |
| Number of runs: | 10 | 9 |
| p-value (runs test): | 0.702 | 0.549 |
| Deviation from linearity? | Not Significant | Not Significant |

The distributions of the flare equivalent durations versus flare rise times are shown for 59 fast and 77 slow flares. Using the Least-Squares Method in the SPSS V17.0 (Green et al., 1999) software, the best fits were predicted for these distributions. The regression computations showed that the best fits are the linear fits given by Equations (3) and (4). All the distributions and the fits are shown in Figure 5.

$$Log(T_r) = 1.262 \times Log(P_B) - 0.793 \quad \text{...................... (3)}$$

$$Log(T_r) = 1.165 \times Log(P_B) - 1.424 \quad \text{..................... (4)}$$

Using the Least-Squares Method in the GrahpPad Prism V5.02 (Dawson & Trapp, 2004), the linear fits were compared whether the separations are real, or not. The results are listed in Table 4.

The slopes of the linear fits given by Equations (3) and (4) are 1.262±0.1094 for the fast flares and 1.165±0.09284 for the slow flares. The *p-probability value* (hereafter *p-value*) was computed to determine whether the difference between the values of the slopes is statistically meaningful. The *p-value* was computed as 0.502. According to this *p-value*, the difference is so small that the linear fits could be accepted as parallel and the value of 1.213 could be used as the value of the associated slope. Moreover, *y-intercept values* were computed to see whether these fits have identical curves, or not. The *y-intercept values* is -0.7928±0.1915 for the fast flares, while it is 1.424±0.1626 for the slow flares. The *p-value* was computed to determine whether the difference between the values of the *y-intercept values* is statistically meaningful. It was found as *p-value* < 0.0001. According to this *p-value*, the difference is so important that there is large difference between the levels of the fit.

Using the *t-Test* analysis, the difference between the mean averages of the equivalent durations was computed and analysed for 59 fast and 77 slow flares to determine the difference between the energies of the slow and fast flares, which is expected, according to Gurzadian (1988). The mean average of the equivalent durations was computed as 1.287±0.198 for the fast flares, while it was 0.495±0.181 for the slow flares. The *p-value* was computed to examine whether the difference of 0.792 between the equivalent durations in the logarithmic scale is statistically meaningful. It was found as *p-value* < 0.0001. According to this *p-value*, the difference is meaningful. The difference of 0.792 in the logarithmic scale is equal to 16.218 s difference between the mean energies of the groups.

The ratios of the flare decay times to the flare rise times were computed for each flare of 59 fast and 77 slow flares. It was seen that the ratios are above the value of 3.5 for all the fast flares, while the ratios are below the value of 3.5. Because of this, we accepted that the value of 3.5 is an indicator value for the classification of the flares. Considering this indicator, we separated all the flares observed in this study into two types as the fast and slow flares. Then, The distributions of the flare equivalent durations versus flare rise times are shown for all 222 flares in Figure 6. We tested whether the linear fits, given by Equations (3) and (4), represent all the data, or not. We derived the linear fits and their the 95% confidence intervals for all 222 flares. As seen from Figure 6, the linear fits given by Equations (3) and (4) are inside the 95% confidence intervals. Thus, it can be statistically assumed that the linear fits given by Equations (3) and (4) can statistically represent all the data of 222 flares. Examining the time-scales of all the observed flares, the maximum rise time was found to 450 s for the fast flares, while it was found to be 2300 s for the slow flares.

## 4. The One-Phase Exponential Association Models of the Distribution of the Flares

Using the method described by Dal & Evren (2010b), the distributions of the flare equivalent durations versus the flare total durations were modelled for each star in the programme in order to examine whether there is any similarity or difference between the behaviours of their flare activities. Using the Least-Squares Method in the SPSS V17.0 (Green et al., 1999) software, the best fits were predicted for these distributions. According to the tests, the *OPEA* (Motulsky, 2007; Spanier & Oldham, 1987) function given by Equation (5) seems to be the best one to model these distributions. Using the Least-Squares Method in the GrahpPad Prism V5.02 (Dawson & Trapp, 2004), the distributions were modelled for each star.

$$y = y_0 + (Plateau - y_0) \times (1 - e^{-k \cdot x}) \quad \ldots \ldots (5)$$

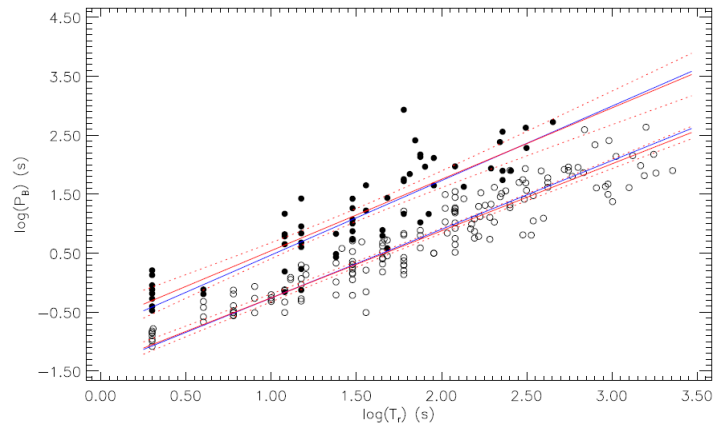

**Figure 6:** The distribution of the equivalent durations versus the flare rise time for 222 flares. In the figure, filled circles represent the fast flares and open circle for the slow flares. The blue lines represent the linear fits shown in Figure 5. The red lines represent the fits derived with using all data of 222 flares. The red doted lines the 95% confidence intervals for each model.

**Table 5:** The parameters derived from the *OPEA* models for each star.

| Star | B-V (mag) | Plateau ($logP_B$) | $y_0$ ($logP_B$) | $k$ | Span Value ($logP_B$) | Half-Life |
|---|---|---|---|---|---|---|
| V1005 Ori | 1.307 | 2.100 ± 0.066 | -0.157 ± 0.139 | 0.00255 ± 0.00034 | 2.256 ± 0.140 | 271.50 |
| AD Leo | 1.498 | 1.887 ± 0.079 | -0.352 ± 0.046 | 0.00240 ± 0.00027 | 2.239 ± 0.084 | 288.80 |
| V1054 Oph | 1.552 | 1.771 ± 0.138 | 0.073 ± 0.139 | 0.00202 ± 0.00061 | 1.698 ± 0.160 | 343.00 |
| EV Lac | 1.554 | 2.548 ± 0.081 | -0.266 ± 0.067 | 0.00227 ± 0.00021 | 2.815 ± 0.090 | 305.70 |
| EQ Peg | 1.574 | 2.997 ± 0.165 | 0.169 ± 0.175 | 0.00122 ± 0.00089 | 2.829 ± 0.144 | 568.40 |

**Table 6:** Using the *t-Test* analysis, the computed mean averages of the flare equivalent durations for the flares, which are located in the *Plateau* phases of the *OPEA* models.

| Star | Mean Average | Std. Deviation |
|---|---|---|
| V1005 Ori | 2.21123 ± 0.08856 | 0.26568 |
| AD Leo | 2.08787 ± 0.15631 | 0.41356 |
| V1054 Oph | 1.73585 ± 0.07880 | 0.15761 |
| EV Lac | 2.65064 ± 0.08066 | 0.19757 |

where, y is the flare equivalent duration in logarithmic scale. x is the flare total duration. $y_0$ is the flare equivalent duration in logarithmic for the minimum flare total duration for each star. The *Plateau value* is the upper limit of the flare equivalent durations in logarithmic scale. According to Equation (2), the *Plateau value* depends on the flare energy. The *Plateau value* must be a saturation level in the flare process occurring on a star. However, $y_0$ depends on both the brightness of the target and the sensitivity of the optic system. *k* is a constant depending on the value of *x*. The models of the distributions are shown in Figure 7 for each star and the parameters derived from the models are listed in Table 5. The *Span value* listed in the table is the difference between $y_0$ and the *Plateau value*. The *Half-Life value* listed in the table is the minimum flare total duration in which the flares reach the saturation level in the flare equivalent duration. Using the *t-Test* analysis, the *Plateau value* derived from each model were tested. The mean averages of the equivalent durations were computed with the *t-Test* for the flares, which are located in the *Plateau* phases of the models. These flares are ones whose flare total times are longer than two times of the *Half-Life values* for each star. The results are listed in Table 6. In the *t-Test* analyses, any error and standard deviation could not be computed for EQ Peg. This is because there is no much more flare in the *Plateau* phase of the model for EQ Peg, and this is why EQ Peg is not listed in Table 6 and Panel (b) of Figure 8.

The stars were compared with respect to their Plateau values. The Plateau values of EV Lac and EQ Peg, which are the reddest stars among all programme stars, are rather higher than others. The mean averages of the flare equivalent durations are also rather higher as expected. The variations of both the Plateau and the mean averages versus B-V indexes of the stars are shown in Figure 8. In the figure, the dotted lines are the linear fits for other three stars, which are shown by filled circles in the figure. Both EV Lac and EQ Peg are shown by open circles. Like the *Plateau* and the mean averages of the flare equivalent durations, the *Span values* of EV Lac and EQ Peg is also rather higher than the same parameters of other stars, which is shown in Figure 9a. This is an expected case. If the $y_0$ parameters are the same for each star in the programme, the *Span values* must be rather higher for EV Lac and EQ Peg, because their *Plateau value* are higher than others. We expect that the $y_0$ parameters are the same for all stars, because all the stars in the programme are almost in the same brightness and all observations were carried out with the same optic system. As it is seen in Table 5, although the $y_0$ parameters seem to vary in a large range, the range of this variation is not as large as the range of the *Plateau* variation.

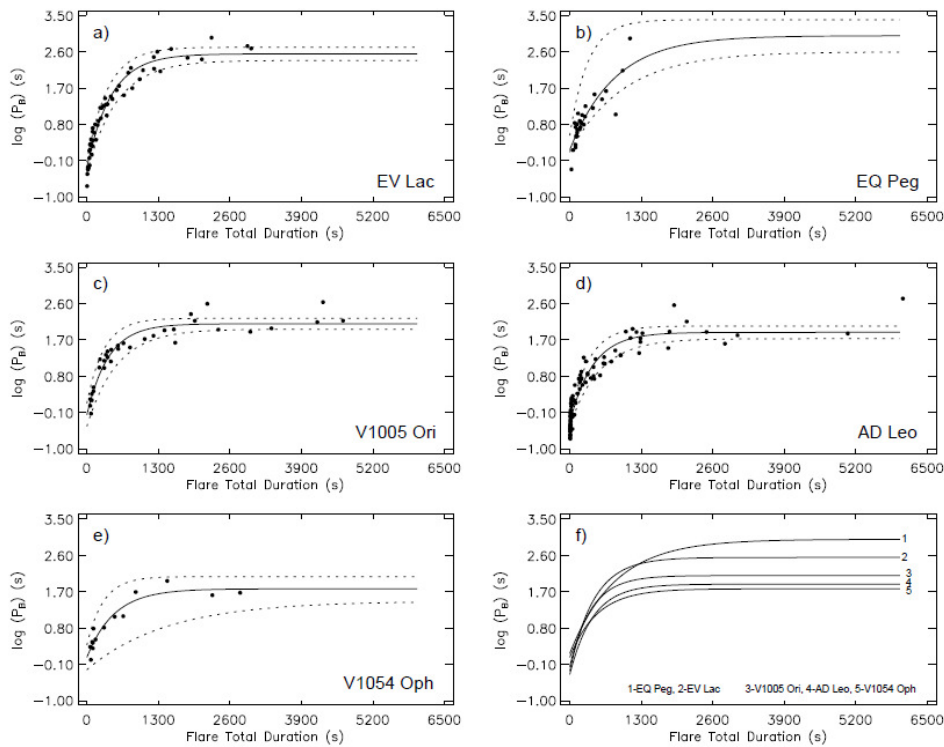

**Figure 7:** The distributions of the flare equivalent durations versus the flare total times for each star (filled circles). In the figure, the lines represent the *OPEA* models. The dotted lines represent the 95% confidence intervals for each model (a, b, c, d, e). In panel f, the model lines are compared.

**Table 7:** Maximum values obtained for the flare rise times and the flare total durations.

| Star | B-V (mag) | Max. Rise Time (s) | Max. Total Duration (s) |
|---|---|---|---|
| V1005 Ori | 1.307 | 2270 | 5561 |
| AD Leo | 1.498 | 1812 | 6053 |
| V1054 Oph | 1.552 | 1260 | 2784 |
| EV Lac | 1.554 | 960 | 2985 |
| EQ Peg | 1.574 | 345 | 1095 |

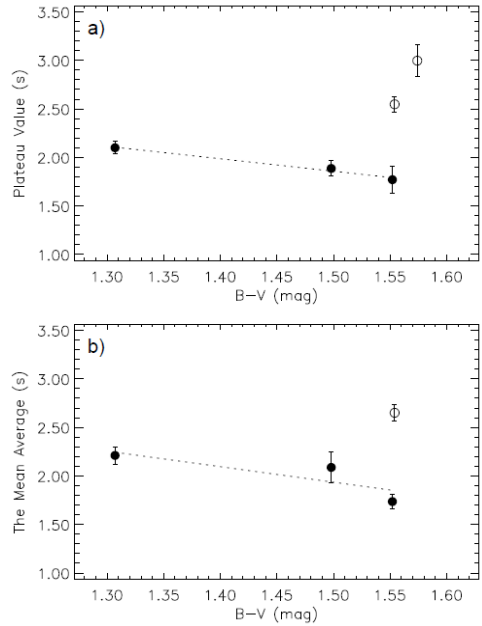

**Figure 8:** The variations of both the *Plateau value* and the mean averages of the flare equivalent durations versus B-V indexes of the stars. The *Plateau values* were derived from the *OPEA* models. The mean averages of the flare equivalent durations were computed by *t-Test* for the flares, which are located in the phase of the *Plateau* in the *OPEA* models. Open circles represent EV lac and EQ Peg. Filled circles represent the other stars in the programme. The dotted lines are the linear fits derived for V1005 Ori, V1054 Oph and AD Leo.

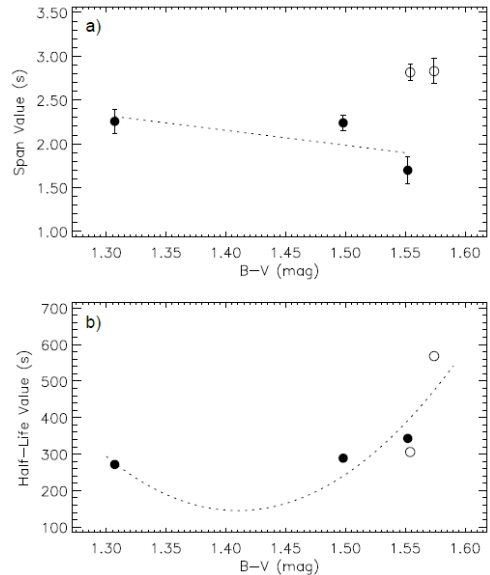

Figure 9: The variations of the *Span values* and the *Half-Life values* versus B-V indexes for the programme stars. Both the *Span values* and the *Half-Life values* were derived from the *OPEA* models. The symbols in Panel a are the same with Figure 8. The doted lines in Panel b are shown to just reveal the variation trend.

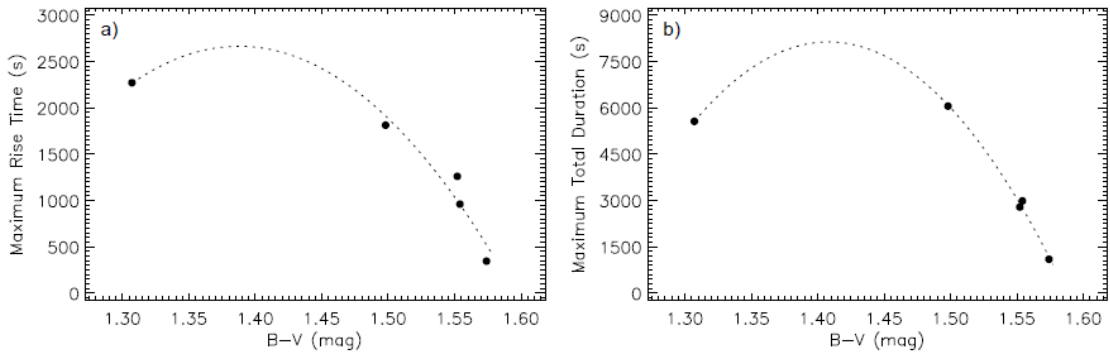

**Figure 10:** The variation of maximum values obtained for flare rise times and flare total duration is shown versus B-V indexes of the stars. Both flare rise time and flare total durations were obtained from the analyses. In the figure, the doted lines represent the polynomial fits in the second degree. The fits are just demonstrations for the way of variations.

Maximum values obtained for the flare rise times and the flare total durations are listed in Table 7 for each star. As it is seen in Figure 10, maximum values of both flare rise times and total durations decrease toward the later spectral types. This means that the lengths of flare phases decrease toward the later spectral types.

## 5. Flare Frequencies

Two flare frequencies were described by Ishida et al. (1991). Using Equations (6) and (7) taken from Ishida et al. (1991), $N_1$ and $N_2$ frequencies were computed for each observing season.

$$N_1 = \Sigma n_f / \Sigma T_t \quad \ldots\ldots\ldots\ldots\ldots\ldots (6)$$

$$N_2 = \Sigma P_B / \Sigma T_t \quad \ldots\ldots\ldots\ldots\ldots\ldots (7)$$

where $\Sigma n_f$ is the total number of the flares observed in a season. $\Sigma T_t$ is the total length of all observing durations in a season. $\Sigma P_B$ is the sum of equivalent duration of all the flares observed in a season. Both $\Sigma T_t$ and $\Sigma P_B$ are in the unit of hours.

All the computed frequencies are listed in Table 8. The $N_1$ frequency of AD Leo is 1.233 for the season 2005-2006. This value is the highest frequency reached among the programme stars. Moreover, The $N_2$ frequency of AD Leo is 0.032 in the same season. And this frequency is also the highest value among the $N_2$ frequencies. According to $N_1$ frequency of the season 2005-2006, AD Leo exhibited more than one flare per hour. According to $N_2$ frequency of the season 2005-2006, the flares occurred on AD Leo had high energy. The second highest flare frequency belongs to V1005 Ori. Both $N_1$ and $N_2$ frequencies of V1005 Ori are high.

## 6. Results and Discussion

In this study, 222 flare were detected in the B band observations carried out from 2004 to 2006. The flares, whose rise times are equal, were identified. Considering their light variations, the identified flares were separated into two groups as fast and slow flares. The distributions of the flare equivalent durations versus the flare rise times were obtained in logarithmic scales and the distributions were fitted by the linear function.

The slopes of the linear fits are 1.262±0.109 for the fast flares and 1.165±0.093 for the slow flares. The *p-value* computed for the slopes demonstrated that the difference between the slopes of the fits is not statistically meaningful. This means that the linear fits are almost parallel. In other words, the equivalent durations for both the fast and slow flares are increasing in the same way with the increasing of the flare rise times. The slopes obtained from the linear fits are almost equal to the slopes obtained for U band flares by Dal & Evren (2010a). The slopes were found as 1.227 for the fast flares and 1.109 for the slow flare by Dal & Evren (2010a). On the other hand, the *y-intercept values* are -0.793±0.192 for the fast flares and -1.424±0.163 for the slow flares. The *p-value* demonstrates that the difference between the *y-intercept values* is statistically meaningful. There is a difference of 0.631 between these two values. Dal & Evren (2010a) found the difference of the values for U band flares as 0.703. This means that the difference between the energy levels of the flare types for U band flares is rather larger than the energy levels of flare types for B band flares. This must be because the energy emitting in U band is larger than that in B band.

Table 8: The flare frequencies of stars for each observing season are listed. In the table, star name and observing seasons are given in first two columns. In the next column, the total observing durations ($\Sigma T_t$) are listed for each season. Then, the total numbers of flares obtained in a season ($\Sigma n_f$) are given. In 5$^{th}$ column, the total equivalent durations obtained from all flares detected in that observing season ($\Sigma P_B$) are listed. In the last two columns, the flare frequencies ($N_1$ and $N_2$) are given.

| Star | Season | $\Sigma T_t$ (h) | $\Sigma n_f$ | $\Sigma P_B$ (h) | $N_1$ (h$^{-1}$) | $N_2$ |
|---|---|---|---|---|---|---|
| V1005 Ori | 2004-2005 | 30.14028 | 5 | 0.05108 | 0.166 | 0.006 |
| | 2005-2006 | 31.40111 | 29 | 0.86013 | 0.924 | 0.029 |
| AD Leo | 2004-2005 | 39.69083 | 36 | 0.44733 | 0.907 | 0.023 |
| | 2005-2006 | 38.91617 | 48 | 0.12379 | 1.233 | 0.032 |
| | 2006-2007 | 24.11472 | 9 | 0.15837 | 0.373 | 0.015 |
| V1054 Oph | 2004 | 18.43750 | - | - | - | - |
| | 2005 | 33.16222 | 14 | 0.07801 | 0.422 | 0.013 |
| EV Lac | 2004 | 6.51111 | - | - | - | - |
| | 2005 | 30.31083 | 22 | 0.08830 | 0.726 | 0.024 |
| | 2006 | 48.76250 | 30 | 0.51510 | 0.615 | 0.013 |
| EQ Peg | 2004 | 9.50972 | - | - | - | - |
| | 2005 | 37.46639 | 29 | 0.34793 | 0.774 | 0.021 |

The *t-Test* analysis (Wall & Jenkins, 2003; Motulsky, 2007; Dawson & Trapp, 2004) was used to determine whether the flare types identified in the analyses are really different from each other. The *t-Test* analysis showed that the mean average of the flare equivalent durations is 1.287±0.198 for the fast flares, while it is 0.495±0.181 for the slow flares. The difference of 0.792 between the mean averages in logarithmic scale is equal to 16.218 times difference between the energies of these two flare types. The *p-value* demonstrates that the difference between the mean averages is statistically meaningful. Gurzadian (1988) stated that there is a large energy difference between the fast and slow flares. The difference of 16.218 in the energies of two flare types must be the difference mentioned by Gurzadian (1988). However, the energy difference between two flare types was found as 157 times for U band flares (Dal & Evren, 2010a). In this case, the energy difference between two type flare is different in U and B bands. In fact, this difference between U and B bands is seen between the estimated B band flare energies given Table 3 and U band flare energies given by Dal & Evren (2010a). This is an expected case according to the results (see Fig.1) given by Gurzadian (1988).

The ratios of the flare decay times to the flare rise times were computed for all observed flares. It was seen that the ratios are above the value of 3.5 for all the fast flares, while the ratios are below the value of 3.5. This means that the value of 3.5 is a limit value between two flare types. Dal & Evren (2010a) also found the limit ratio is 3.5 for U band flares. Therefore, we accepted that the value of 3.5 is an indicator value for the classification of the flares. Considering this indicator, we separated all the flares observed in this study into two types as the fast and slow flares. A similar classification is also exist for the Solar flares (Temmer et al., 2001). Concerning the statistical analysis of solar H$\alpha$ flares, the author describe the case as the asymmetry of the decay and rise times instead of the term of the fast or slow flares.

Providing that the value 3.5 is a limit ratio for the flare types, the slow flare rate is 73% of all 222 flares, while the fast flare rate is 27%. According to Dal & Evren (2010a), it is 37% for the U band fast flares, and it is 63% for the U band slow flares. There is a difference of the flare type rates between U and B band flares. This must be because the slow flares exhibit themselves in B band more obviously, while the fast flares exhibit themselves more obviously in U band. This is because the fast flares are more energetic events than the slow flares.

The longest flare rise time is 450 s for the fast flares, while it can reach 2300 s for the slow flares. It is seen that the length of the rise times can reach 400 s for the fast flares and 1400 s for the slow flares for U Band flares (Dal & Evren, 2010a). As it is seen the maximum flare rise times are longer in B band than in U band. This could be because the B band is more sensitive than U band for cooler structures on the surfaces of the stars. In this case, although it is seen that the brightness during a flare decreased completely in the quiescent level in U band, it is seen that the B band brightness has not completely decreased until the quiescent level. This case could be changeable for the fast and slow flares. If both the flare energies and the numbers of observed flares are considered, it can stated that U band is more sensitive to observe and examine the fast flares, while B band is more sensitive to observe and examine the slow flares.

Moreover, the value of 3.5, the ratio of flare decay times to flare rise times, must be a critical value between the thermal and non-thermal processes. The value of the ratio can give an idea about the rate of energy emitting in flare processes. In this study, obtained maximum flare rise time is 450 s for the fast flares, while it is 2300 s for the slow flares. On the other hand, there is no limit length for the maximum decay times of both flare types. In this case, the ratio of flare decay times to flare rise times depends on the flare rise time more than the flare decay time. The flare rise time is a duration length computed from the phase, in which the energy is suddenly released. In this phase, some region(s) on the surface of the star is heating and consequently, the brightness of the star increases. If the region is rapidly heating, the flare rise time will be shorter. In conclusion, the flare rise time depends on the rate of heating on the star surface. This is why the ratio of flare decay times to flare rise times could be a critical parameter for the thermal and non-thermal processes.

Besides, the flare time-scales reveal some properties of the flaring loop geometry (Reeves & Warren, 2002; Imanishi et al., 2003; Favata et al., 2005; Pandey & Singh, 2008). Especially, Favata et al. (2005) demonstrated that that much longer flares are arcade flares, while short flares generally occur in a single loop. As seem from Table 7 and Figure 10, the flare time-scales are dramatically decreasing toward the later spectral types. The interesting case is that the maximum flare rise times behave in the same trend with the maximum flare total times. This indicates that a systematic phenomenon in the flaring loop geometry occurs toward the later spectral types. However, as seen from Table 5 and Figure 9, the *Half-Life* parameter increases toward the later spectral types. The *Half-Life* parameter is a specific flare total durations, for which the models reach the saturation levels. Consequently, the time-scales of the flare event are decreasing toward the later spectral types, while the times needed to reach the saturation level in the energy are increasing.

The distributions of the flare equivalent durations versus the flare total durations were ob tained for five programme stars. Using the *OPEA* function, all the distributions were modelled. According to the models, the flare equivalent durations become stable in a specific level after a specific flare total duration. The specific level of the flare equivalent durations is called the *Plateau values* in the models. The *Plateau value* is different for each star. The *Plateau values* of both EV Lac and EQ Peg, which are the reddest stars in the programme, are especially higher than other programme stars. In a sense, the *Plateau value* is a saturation level of the equivalent durations for each star. As it is seen in Figure 7, the saturation level of the stars is different from each other. The *t-Test* analysis was used to test whether the *Plateau values* show truly maximum level for equivalent durations. As it is seen in Figure 8, the mean averages computed by *t-Test* analysis and listed in Table 6 are in agreement with the *Plateau values*. Both the *Plateau values* and the mean averages, gradually decrease toward the later spectral types for V1005 Ori, AD Leo and V1054 Oph, while these parameters are dramatically higher for the reddest stars, EV Lac and EQ Peg.

If the flares with high energy are an indicator of high flare activity level for a star and if the high flare activity level depends on the age of a star, according to Skumanich (1972), we expect that the flares with high energy should have been detected in the observations of V1005 Ori. This is because V1005 Ori is a member of the 35 million-year old IC 2391 Supercluster. The star is the youngest target in the programme. On the other hand, considering that all programme stars are almost at the same ages apart from V1054 Oph, the differences among the *Plateau values* seem to be caused by different equatorial rotational velocities (Veeder, 1974; Fleming et al., 1995; Montes et al., 2001). As seen from the chromospherically active stars, assuming that the Skumanich law is also acceptable for the flare activity, we expect that the flares with high energy should have been detected from the stars with high rotational velocity. The equatorial rotational velocity of EV Lac is 4 $kms^{-1}$ and it is 5-5.8 $kms^{-1}$ for AD Leo (Marcy & Chen, 1992; Pettersen, 1991). However, the equatorial rotational velocity of V1005 Ori is 29.6 $kms^{-1}$ (Eker et al., 2008). In conclusion, the most active flare star should have been V1005 Ori. If the high flare activity level depends on binarity or multiplicity more than other physical parameters of the stars, the flare activity levels of both EQ Peg and especially V1054 Oph must be higher than other stars. EV Lac, AD Leo and V1005 Ori are known as the single stars. Furthermore, EQ Peg has a visual companion (Pettersen, 1991). Moreover, V1054 Oph is a member of the system with six companions, and the star is a triple system itself (Pettersen et al., 1984; Mazeh et al., 2001). Therefore, the flares detected in the observations of V1054 Oph should have been the most energetic flares. It must be remembered that the flare stars are known as the young stars, especially pre-main sequence stars (Mirzoyan, 1990; Gershberg, 2005; Benz & Güdel, 2010). However, there are stars in close binary systems which are very old, but have a very high level of flare activity (Poveda et al., 1996a,b; Rocha-Pinto et al., 2002). This is because such systems are tidally locked, so that the two stars in a binary system can preserve their high rotational velocities instead of spinning down with age like single stars

The X-Ray observations of the solar flares demonstrated that some parameters in the process lead to the energy emitting (Gershberg, 2005). The suspected parameters are $\upsilon_A$, *B*, *R* or $n_e$. The Alfv´en velocity ($\upsilon_A$) is proportional to the magnetic field *B*. The emissivity of the plasma *R* depends on the electron density ($n_e$) of the plasma (Van den Oord & Barstow, 1988; Van den Oord et al., 1996). The suspected parameters must be *B* or/and $n_e$ in the flaring loop. According to Katsova et al. (1987), the most effective parameter is the electron density $n_e$ in the volume of the loop. In the case of the white-light stellar flares, these parameters must also be effective on the energy emitting in a flare.

If the flares with high energy are an indicator of high flare activity level, according to both the *OPEA* models and the *t-Test* analyses, the most active stars are EV Lac and EQ Peg. On the other hand, according to flare frequencies, AD Leo and V1005 Ori are the most active stars among all others. For instance, according to the $N_1$ frequency, the number of the flares detected in the observations of AD Leo in the observing season 2005-2006 is 1.233 per hour. It is 0.904 per hour for V1005 Ori in the same season. According to $N_2$ the frequencies of these stars, each flare is very highly energetic. In contrast to the results obtained from both the *OPEA* models and the *t-Test* analyses, EV Lac and EQ Peg exhibited low level flare activity according to the $N_1$ frequency. Their flares are also low energetic according to their $N_2$ frequency. However, Leto et al. (1997) showed that the flare frequency of EV Lac used to increase from 1967 to 1977. Comparing the flare frequency together with the annual averages of the brightness in B band of EV Lac, Mavridis & Avgoloupis (1986) demonstrated that EV Lac exhibits a flare activity cycle of 5 years. Considering these studies, if both EV Lac and EQ Peg exhibit a flare activity cycle, the stars could be in the minimum term of the activity cycles. This can explain that both the *Plateau values* and the mean averages are dramatically higher than other stars, while the flare frequencies are not higher.

On the other hand, according to some observational results, taking the flare frequency as an indicator of the flare activity level is not trustable to determine the flare activity level. For instance, the flare frequency of V1005 Ori was found to be 0.904 per hour in the season 2005-2006, while it was found to be 0.166 per hour in the next season in this study. The same cases are seen for other UV Ceti type stars in the literature (see Nicastro 1975; Contadakis et al. 1982). In addition, Leto et al. (1997) revealed an increasing in the flare frequency of EV Lac during 10 years, and Mavridis & Avgoloupis (1986) demonstrated a flare activity cycle of

5 years for EV Lac. However, considering the large data set obtained along 18 years, Ishida et al. (1991) found no variation in the flare frequencies ($N_1$ and $N_2$) of EV Lac. According to us, this is because that the computed flare frequency depends on the total observing duration in a season as well as the flare activity levels of the stars.

In conclusion, we analysed and modelled the flare data obtained from the B band observations of five stars. Six basic properties have been found from all models and analyses. 1) According to the method described by Dal & Evren (2010a), the flares were separated into two types as fast and slow flares. It was seen that the value of 3.5 derived from the ratio of flare decay time to flare rise time is an indicator. This value is equal to the value found by Dal & Evren (2010a). 2) There is a 16.2±3.7 times difference between the averages of the equivalent durations of slow and fast flares. 3) It is no matter how long the flare total durations, maximum flare energy is below the specific level of energy for each star. The level given as the *Plateau value* in the models is changing from star to star. This *Plateau value* in the models must be the saturation level for the flare processes occurring on the stars. 4) The *Plateau values* of EV Lac and EQ Peg, which are the reddest stars among the programme stars, are absolutely higher than the others. 5) Minimum values of the flare total durations, which are needed to reach the saturation level of the flare energy, increase toward the later spectral types. This value is called *Half-Life value* in models. 6) Both maximum flare rise times and flare total durations obtained from the observed flares are decreasing toward the later spectral type.

Finally the results obtained from B band flare data are in agreement with the results obtained from U band flares. This result demonstrates that the methods described by Dal & Evren (2010a,b) are statistically useful and trustworthy for the statistical analyses of flare data sets.

**Acknowledgments**

The authors acknowledge generous allotments of observing time at Ege University Observatory. We thank the referee for useful comments that have contributed to the improvement of the paper. We wish to thank Prof. Dr. M. Can Akan gave us valuable suggestions, which improved the language of the manuscript. We also wish to thank the Ege University Research Found Council for supporting this work through grant Nr. 2005/FEN/051.